\documentclass[aps,preprint]{revtex4}
\usepackage[dvipsnames]{xcolor}
\usepackage{amsfonts}
\usepackage{amssymb}
\usepackage{subcaption}
\usepackage{amsmath}
\usepackage{subcaption}
\usepackage{graphicx}
\usepackage[font={footnotesize,it}]{caption}
\usepackage{bigints}
\usepackage{latexsym}
\usepackage{enumitem}

\begin{document}

\title{Colliding null matter with a specific stress tensor}% Force line breaks with \\
%%\thanks{A footnote to the article title}%

\author{Mustafa Halilsoy}
 \email{mustafa.halilsoy@emu.edu.tr}

\author{Chia-Li Hsieh}%
 \email{galise@gmail.com}

\author{Mert Mangut}%
 \email{mert.mangut@emu.edu.tr}

\affiliation{%
 Department of Physics, Faculty of Arts and Sciences, Eastern Mediterranean University, Famagusta,
North Cyprus via Mersin 10, Turkey %%\textbackslash\textbackslash
}%

\begin{abstract}
The accretion disks around black holes consist of infalling matter boosted almost to the speed of light making collisions with opposite counterpart. This is the rough picture occurring near black holes or other strongly gravitating centers that produce observed phenomena such as astrophysical jets. A toy model that can be considered imitating such a process is colliding null sources in general relativity. We present such a simple model projected into the plane of null coordinates that takes into account only neutral sources. We show that even at such a simplified model, uncharged and non-rotating, it is possible to obtain jet-like ejections albeit they lie below the horizon. In the present study the spacetime consists of either one of i) a cloud of strings, ii) a global monopole, iii) a particular model of bumblebee gravity, all described by a similar class of stress-energy tensor. There are gravitational waves accompanying the null sources and naturally collision of gravitational waves is also taken into account. After the collision, the spacetime contains both null and non-null sources, followed by trailing gravitational radiations. Locally the interaction region of the colliding null-sources and gravitational waves is isometric to the static background spacetime.
%%\begin{description}
%%\item[Usage]
%%Secondary publications and information retrieval purposes.
%%\item[Structure]
%%You may use the \texttt{description} environment to structure your abstract;
%%use the optional argument of the \verb+\item+ command to give the category of each item.
%%\end{description}
\end{abstract}

%Use showkeys class option if keyword
                              %display desired
\maketitle

%\tableofcontents

%%==================================================================================================

\section{\label{sec:level1}Introduction}

In general relativity (GR), a static metric mostly has a dynamical region in which upon a coordinate transformation we can introduce waves moving in opposite directions. The existence of such a region renders it possible to define plane waves as substructures in the underlying spacetime and their collision. We recall that the inner horizon region of a Schwarzchild black hole turns dynamical whereas outside horizon region remains static. As a matter of fact, this behavior is valid for all Schwarzchild-like black hole solutions in any theory of gravity.

Plane wave spacetimes are expressed either in Brinkmann [1] or Rosen forms [2]. The former is expressed in a single-null coordinate whereas the latter admits double-null coordinates and becomes more apt for the problem of collision [3]. We recall that due to high symmetry, admitting 5-Killing vectors, the plane wave spacetimes constitute the simplest class of spacetimes in GR. We also remind that whereas there exists gravitational interaction between sources of any sort, it must be provided through exchanges of gravitational waves and plane waves are the simplest intermediating agents to handle. It is known that a gravitational wave is the classical version of the graviton which is to be considered in the quantum domain. What we observe and detect globally throughout our devices are at the classical level so that GR is the right address for it.

A null source can be considered as a static source with an infinite boost such that the observers see a source moving at the speed of light. Null-shells constitute such matters that have been considered in the past extensively. The collision of such shells with Dirac delta sources has been studied long ago \cite{4,5,6}. In GR it is known that any moving source creates its own gravitational waves. These may be chock or impulsive types which are quite strong to self-focus itself together with the accompanying source. The nonlinearity of GR makes these focusing effect strong enough to modify all participating sources and create undesired singularities. The creation of null and nonnull energy momenta components reflects the nonlinear manifestation of GR.

In this paper, we consider a particular source described as the asymptotic form of a global monopole \cite{7} in static form or a cloud of strings \cite{x}. At the same time, this Schwarzchild-like metric that provides a solution  to the bumblebee gravity \cite{8}. By transforming such a static metric to the space of double-null coordinates, we can study the collision of boosted sources. We remind that in analogy with a point electric charge that transforms into a plane electromagnetic wave under an infinite Lorentz boost, a static mass transforms into a gravitational wave \cite{9}. The yields of such a violent collision process are, besides the excess flux of gravitational waves, new components of energy-momentum. Prior to the collision, we have a null source of matter accompanied with a Weyl curvature component, whereas after the collision we have additional components. In general, resulting sources consist of two types: i) impulsive (with a delta function) type, and ii) shock ( with a Heaviside step function) type. The local isometry between spacetimes admitting 2- spacelike and 1- spacelike, 1- timelike Killing vectors was suggested first by Chandrasekhar and Xanthopoulos (CX) \cite{10}.

Such a collision process can be an appropriate mathematical toy model to describe the astrophysical jets encountered at the central regions of strongly attractive sources, such as black holes (see \cite{11,19,20,21} for a review of the subject). Namely, the sources attracted by a central black hole reaches almost to the speed of light and collides with oppositely moving sources to give as outcome, the jets observed in astrophysics. In the present paper, we consider only collision of linearly polarized sources and waves. Expectedly, if the incoming sources/waves possess additional degree of polarization, the outcoming sources will be spinning. Another physical factor that we didn’t take into account is the possibility of sources with electric/magnetic fields. We are confined only to the case of colliding neutral matter boosted to the speed of light which yields impulsive and shock sources accompanied with gravitational waves. For such reasons our model can be considered at best as a toy model. We investigate the prototype of a shock source that emerges after collision in some details. Depending on the difference in strengths of the fluxes prior to the collision, we have symmetric localization of energy outcome. The simplest possible outcome corresponds to the case of symmetric collision where from both sides sources income with equal energies. Different energy parameters implies asymmetrical collisions and outcomes.

We note finally that our considerations cover only the projection of the process into a two-dimensional plane. That is, in the null coordinates ($u, v$) or equivalently ($t ,z$) coordinate in 1-space and 1-time variables. The entire process is to be understood as projected into the plane of these two-coordinates. The remaining two coordinates, say ($x, y$) are to be understood as orthogonal to the plane of ($u ,v$) or ($t ,z$). Unfortunately, this is the price to be paid when we solve the problem exactly within the capacity of GR. The occurrence of two space-like Killing vectors throughout spacetime before and after the collision brings us this restriction. To our knowledge in the literature of GR and astrophysics there is no work investigating the jets from the collision point of view. The absence of such an approach motivated us to make such a research. Starting from chargeless and spinless sources may sound simple but yet nontrivial. Once this is completed our next venture will be to consider charged, spinning sources making collision in the vicinity of BHs. This will give rise to spinning jets and also possible $\gamma-$ ray bursts as yields from the collision of highly boosted charges. 

The paper is organized as follows. In section II, we introduce our static spacetime, representing the source and its transformation to the double-null coordinates. In section III, we formulate the collision problem of our null-source accompanied with gravitational waves. We interprete our spacetime of colliding null matter in section IV. The paper is completed in section V with our conclusion. The sources and Weyl curvature are tabulated in Appendix A.

%%===============================================================================

%%%%%%%%%%%%%%%%%%%%%%%%%%%%%%%%%%%%%%%%%%%%%%%%%%%%%%%%%%%%%%%%%%%%%%%%%%%%%%%%%

%%%%%%%%%%%%%%%%%%%%%%%%%%%%%%%%%%%%%%%%%%%%%%%%%%%%%%%%%%%%%%%%%%%%%%%%%%%%%%%%%%%%
%%==================================================

\section{The Static Spacetime}

We start with the static metric described by the line element
\begin{equation}
  ds^2=(1-2k-\frac{2m}{r})dt^2-\frac{dr^2}{(1-2k-\frac{2m}{r})}-r^2(d\theta^2+\sin^2\theta d\phi^2),
\end{equation}
which represents the asymptotic form of a global monopole  \cite{7}. The constant parameter $k$ characterizes the strength of the monopole. For $k=0$, we recover the vacuum Schwarzchild metric. The energy-momentum tensor $T_\mu^\nu$, is given from the Einstein equations
\begin{equation}
G_\mu^\nu = T_\mu^\nu =
\begin{pmatrix}
\frac{-2k}{r^2} & 0 & 0 & 0 \\
0 & \frac{-2k}{r^2} & 0 & 0 \\
0 & 0 & 0 &0 \\
0 & 0 & 0 &0
\end{pmatrix}
\end{equation}
which suggests that the spacetime represents also a cloud of strings \cite{x} . We remark that a similar Schwarzchild-like static metric has been given in a similar form. Namely, the metric given by

\begin{equation}
  ds^2=(1-\frac{2M}{r})dt^2-(1+l)\frac{dr^2}{(1-\frac{2M}{r})}-r^2(d\theta^2+\sin^2\theta d\phi^2),
\end{equation}

where again, $M$=mass and $l$=constant, the Lorentz symmetry breaking constant, solves the static bumblebee gravity equations \cite{8}. In other words, all the three different cases referred to possess the same type of energy-momentum tensor (2) which will be considered in the paper. It can be easily checked that up to a scale transformation of time, we have in the foregoing metrics (1) and (3), the relations

\begin{equation}
  M=\frac{m}{1-2k},
\end{equation}

\begin{equation}
  l=\frac{2k}{1-2k},
\end{equation}
which make the two metrics identical. Once we show that these are identical, we choose (1) also as the representative of (3). We apply a coordinate transformation on (1) to transform it into the double-null coordinate form so that we can formulate the problem of collision of sources. By source clearly, it is to be understood the boosted form of the static metric in null-coordinates.

There is no loss of generality in making the choice for mass
\begin{equation}
  m=1-2k,
\end{equation}
and apply the coordinate transformation
\begin{equation}
  t=\frac{\sqrt{2}x}{1-2k},
\end{equation}

\begin{equation}
  \phi=\frac{\sqrt{2}y}{ \sqrt{1-2k}},
\end{equation}

\begin{equation}
  r=1+\sin(u+v),
\end{equation}

\begin{equation}
  \theta=\frac{1}{\sqrt{1-2k}}(u-v)-\frac{\pi}{2}.
\end{equation}

Upon rescaling
\begin{equation}
  ds^2\rightarrow\frac{1-2k}{2}ds^2,
\end{equation}
we obtain the line element in the form
\begin{equation}
  ds^2=X^2(2dudv-\cos^2(\frac{u-v}{\sqrt{1-2k}})dy^2)-\frac{\Delta}{X^2}dx^2,
\end{equation}
where we abbreviated

\begin{equation}
  X=1+\sin(u+v),
\end{equation}
and
\begin{equation}
  \Delta=\cos^2(u+v).
\end{equation}

This line element is our basic tool that we shall concentrate in this study by interpreting it as a collision geometry for appropriate sources. To see the topology of the incoming waves we suppress one of the null coordinate (say $v$) and obtain from Eq.s (9) and (10) the relation

\begin{equation}
r(\theta)=1+sin\left(\sqrt{1-2k}\left(\theta+\frac{\pi}{2}\right) \right).
\end{equation}

\begin{figure}[h]
\centering
\includegraphics{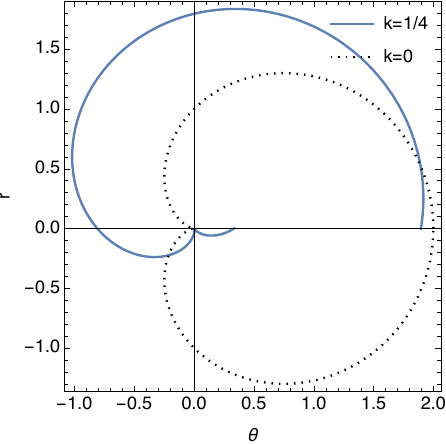}
\caption{The plot of Eq.(15) in polar coordinates for $k=0$ and $k=1/4$ in the domain $0\leq\theta\leq2\pi$. The case $k=0$ is the heart-like shape known as cardioid. The case $k=1/4$ gives rise to a spiral-type curve as the profile of the wavefront. Note that the restrictions on null coordinates limit the ranges of the angular variable $\theta$.}
\end{figure}

\newpage

For pure gravitational wave ($k=0$) it reduces to

\begin{equation}
r(\theta)=1+cos(\theta)
\end{equation}

which, for $0\leq\theta\leq2\pi$, known as a 'cardioid', due to its heart-like shape (Fig.1). The spiral structure of the particular $k=1/4$ is also shown in Fig.1 We note that $\theta-$domain in (15) depends on the source parameter $k<1/2$, as $r(\theta)$ will determine the spacelike singularity. Note also that plotting $r(\theta)$ as the $y-$axis versus $\theta$ as the $x-$axis shows the  shrinking of horizon from $r=2$ to $r=0$, for the case of cardioid.

It should also be added that under the transformation used the parameter $k$ is restricted as
\begin{equation}
0<k<\frac{1}{2},
\end{equation}
or for the corresponding Lorentz symmetry breaking factor

 \begin{equation}
l>0.
\end{equation}

Once we cast our static metric (1) (or (3)) into the double-null coordinate form, we are now ready to discuss the problem of collision.

%%=====================================================================================

\section{Collision of Null-matter Sources}

We insert into the global metric (12) the Penrose substitutions
 \begin{equation}
u\rightarrow u\theta(u),
v\rightarrow v\theta(v),
\end{equation}
where $\theta(u)$ and $\theta(v)$ are the Heaviside step functions satisfying the definition

\begin{equation}
\theta(x) =
\left\{
    \begin{array}{lr}
        1, \qquad   \text{if } &  x > 0\\
        0, \qquad   \text{if } &  x \le 0
    \end{array}
    \right\}
\end{equation}
We also make scaling in the null coordinates as
 \begin{equation}
u\rightarrow au,
v\rightarrow bv,
\end{equation}
where $(a, b)$ are constants with the units of inverse length in the geometrical units. In physical units $(a, b)$ measure energy and for that reason we use this freedom to incorporate the strength of energy of the incoming sources/waves. After the substitutions of (19) and (21) into our line element (12), we choose a null tetrad of Newman-Penrose (NP) \cite{12} to compute the Ricci and curvature components. Our choices is

\begin{equation}
    \begin{array}{lr}
         l_\mu=X(u, v)\delta_\mu^u,\\
         n_\mu=X(u, v)\delta_\mu^v,\\
         m_\mu=\frac{1}{\sqrt{2} X(u, v)}\cos(au\theta(u)+bv\theta(v))\delta_\mu^x \\
         +iX(u, v)\cos(\frac{au\theta(u)-bv\theta(v)}{\sqrt{1-2k}})\delta_\mu^y,
    \end{array}
\end{equation}

and the complex conjugate of $m_\mu$. We recall that our $X(u, v)$ is $X(u, v)=1+\sin(au\theta(u)+bv\theta(v))$.
In this choice of NP tetrad, the non-zero Ricci ($\Phi_{AB}$) and Weyl components $\Psi_A$ are tabulated in Appendix A. By inspecting these components, we can locate the singularities in the metric and obtain the exact profiles of the incoming waves/sources upon substitutions of the step functions. We formulate first the problem of collision as follows. We divide the spacetime into four regions as (Fig.2)

\begin{figure}
  \centering
  \includegraphics[width=1\textwidth]{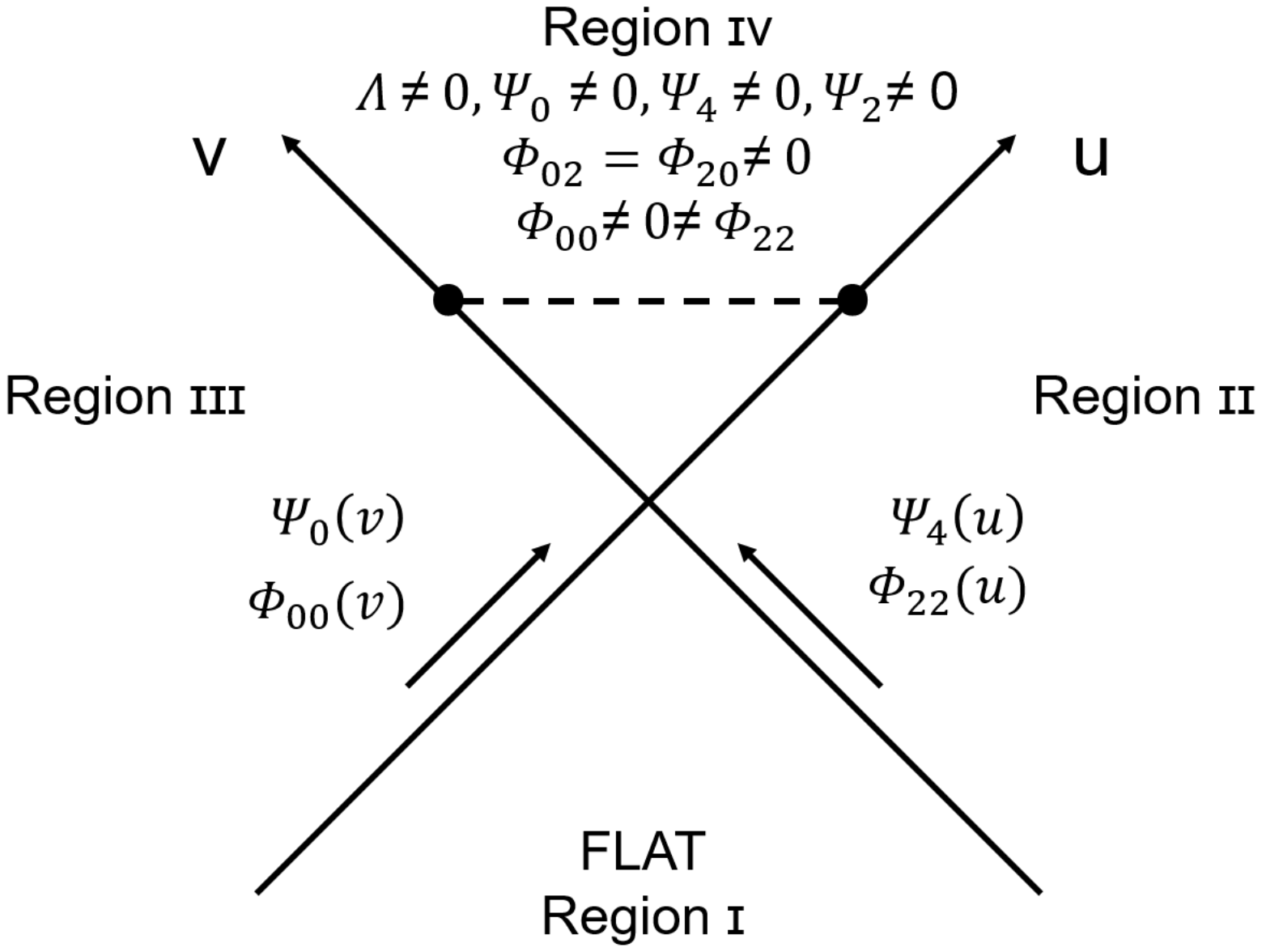}
  \caption{Collision of sources $\Phi_{22}(u)$ and $\Phi_{00}(v)$ coupled with gravitational waves. There are null singularities at ($u=0, v=\frac{\pi}{2b\sqrt{1-2k}}$) and ($v=0, u=\frac{\pi}{2a\sqrt{1-2k}}$), shown as heavy dots on the null coordinates, where $\Psi_4$ and $\Psi_0$ diverge. The sources $\phi_{22}(u)$ and $\phi_{00}(v)$ also become divergent at these points. For the interaction region with $u>0, v>0$, however, the delta function $\delta(u)$ and $\delta(v)$ are not effective and we search for the other terms. We note that in the interaction region, we must impose the condition $au+bv<\frac{\pi}{2}\sqrt{1-2k}$, which otherwise confronts with the 'fold' singularity.}
\end{figure}
\newpage

\subsubsection{Region $I$, $(u<0, v<0)$}

This is the flat region with no matter and no gravitational waves.

\subsubsection{Region $II$, $(u>0, v<0)$}

This is the incoming region from ($+z$), where we define coordinates ($t, z$) as
\begin{equation}
 \sqrt{2}u=t+z, \sqrt{2}v=t-z
\end{equation}
The line element for region II is
\begin{equation}
  ds^2=X^2(u)(2dudv-\cos^2(\frac{au}{\sqrt{1-2k}})dy^2)-\frac{1-\sin(au)}{1+\sin(au)}dx^2,
\end{equation}
where $X(u)=1+\sin(au)$ . It is seen that for a nonzero $g_{yy}$, we must have $au<\frac{\pi}{2}\sqrt{1-2k}$

\subsubsection{Region $III$, $(u<0, v>0)$}
This is same as region II with $u\leftrightarrow v$ and $a\leftrightarrow b$. It represents our incoming source from ($-z$). Similar to region II, we must have $bv<\frac{\pi}{2}\sqrt{1-2k}$.

\subsubsection{Region $IV$, $(u>0, v>0)$}

This represents the interaction (or collision) region given by the line element
\begin{equation}
  ds^2=X^2(u,v)(2dudv-\cos^2(\frac{au-bv}{\sqrt{1-2k}})dy^2)-\frac{1-\sin(au+bv)}{1+\sin(au+bv)}dx^2,
\end{equation}
where $X(u, v)=1+\sin(au+bv)$ and we must have the restriction  $au-bv<\frac{\pi}{2}\sqrt{1-2k}$, which follows from the conditions of the incoming regions. We note that due to the restriction $au<\frac{\pi}{2}\sqrt{1-2k}$ in Region II the metric function $g_{xx}\neq0.$ This is from the fact that $au=\pi/2$ is not accessible. As a matter of fact for $au=\frac{\pi}{2}\sqrt{1-2k}$ we have $g_{yy}=0,$ which corresponds to a coordinate singularity in the Region II. In the collision problem on the other hand due to the singularity at $v=0$, the singularity $au=\frac{\pi}{2}\sqrt{1-2k}$ is called a "fold" singularity. At such a singularity the norm of the Killing vector $\partial_y$ vanishes and curvature diverges. Similar discussion is valid also for the incoming Region III at $u=0,$ $bv=\frac{\pi}{2}\sqrt{1-2k}$. Because of these "fold" singularities we stay on the regular part in Region IV by choosing our domain $au+bv<\frac{\pi}{2}\sqrt{1-2k}.$ The choice $au+bv=\pi\sqrt{1-2k}$ will cause a region in which the signature of the metric component $g_{xx}$ turns sign and therefore we must avoid. For a general discussion of singularities in GR one should consult \cite{16}. For a physical discussion of "fold" singularities in the colliding wave problem we suggest \cite{3} and \cite{18} .

The Ricci and Weyl components in the Appendix reveal about the singularities of the solution. The occurrence of  delta functions concerns about the singularities on the null cone. When we consider the interaction region alone for $u>0, v>0$, the null singularities are discarded. In the interaction region, the indicative term is the one that is included in all the tetrad scalars, namely the combination $\sim\frac{1}{X^2}$ for Ricci’s and $\frac{1}{X^3}$ for Weyl’s scalars. We investigate the combination $\Phi_{22}\sim\frac{1}{X^2}$ in details which describes the energy distribution after the collision. The function $X(u, v)$ involves the $\sin(au+bv)$, which is periodic, however, due to the restrictions on $u$ and $v$ access to $\sin(au+bv)=-1$, is not possible.  With those restrictions we have for the function  $au+bv<\frac{\pi}{\sqrt{2}}\sqrt{1-2k}$ or equivalently $(a+b)t+(a-b)z<\frac{\pi}{\sqrt{2}}\sqrt{1-2k}$. We plot the details of the function $\frac{1}{X^2}$ in both the null $(u, v)$ and the $(t, z)$ coordinates to see the evolution of the Ricci component after the collision.

Our first observation about the Ricci component $\Phi_{22}$ (and $\Phi_{00}$) is that the coefficient of the delta function, namely the function $F(u)=\tan(au)-\frac{1}{\sqrt{1-2k}}\tan(\frac{au}{\sqrt{1-2k}})$ and similarly $G(v)$ with $a\leftrightarrow b$ and $u\leftrightarrow v$, vanish in the incoming regions. This simply means that in the entire allowable domain $au<\frac{\pi}{2}\sqrt{1-2k}$ and $bu<\frac{\pi}{2}\sqrt{1-2k}$  the delta function terms are created and survive on the null boundary. Figs. (3) and (4) display the relevant Ricci components of colliding sources after collision in different coordinate systems.

\newpage

\begin{figure}[h]
\centering
  \begin{tabular}{@{}cccc@{}}
    \includegraphics[width=0.5\textwidth]{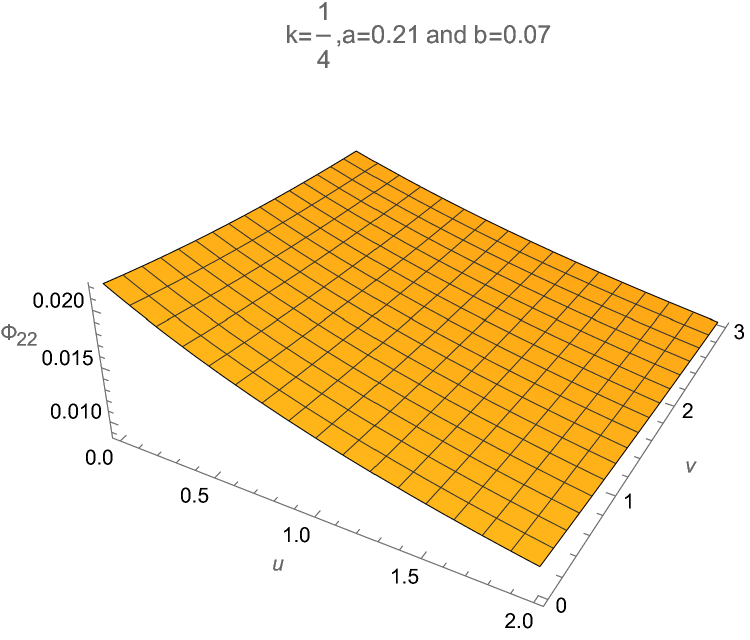} &
    \includegraphics[width=0.5\textwidth]{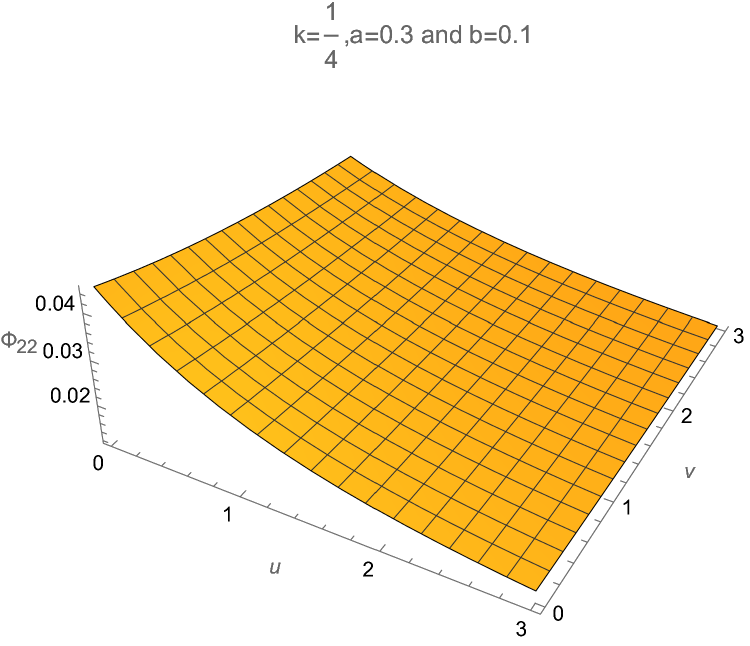} &

                                                             \\
    \multicolumn{2}{c}{\includegraphics[width=0.5\textwidth]{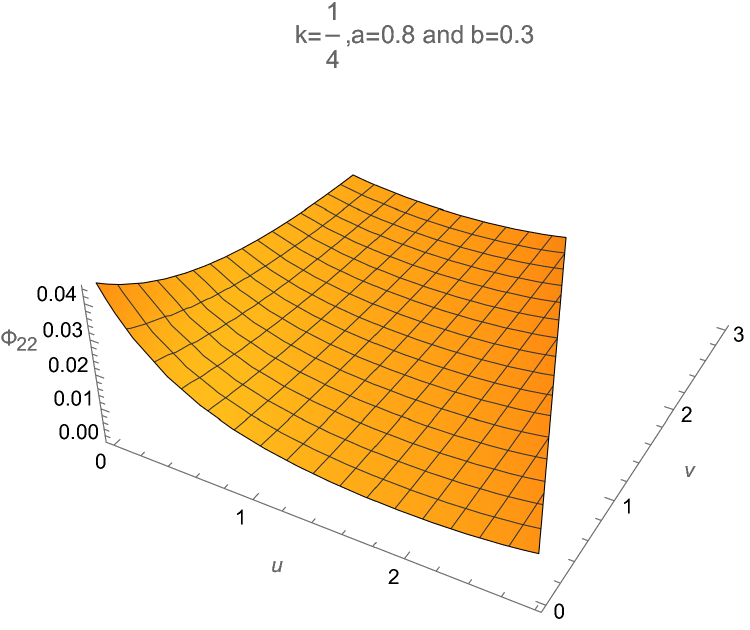}}
  \end{tabular}
  \caption{Post collision plot of the Ricci component $\Phi_{22}=\frac{ka^{2}}{(1-2k)X^{2}}$, for $u>0$, $v>0$ and $k=1/4$, with $X=1+sin(au+bv)$ satisfying the constraint $au+bv<\frac{\pi}{2}\sqrt{1-2k}$. Note that the plots of  $\Phi_{00}(u,v)$ and $\Phi_{02}(u,v)$ also will follow similar behaviour in which starting from the collision point $u=v=0,$ the Ricci components decrease gradually.}
\end{figure}

\begin{figure}[!ht] 
\centering
\begin{subfigure}{0.45\textwidth}
    \includegraphics[width=\textwidth]{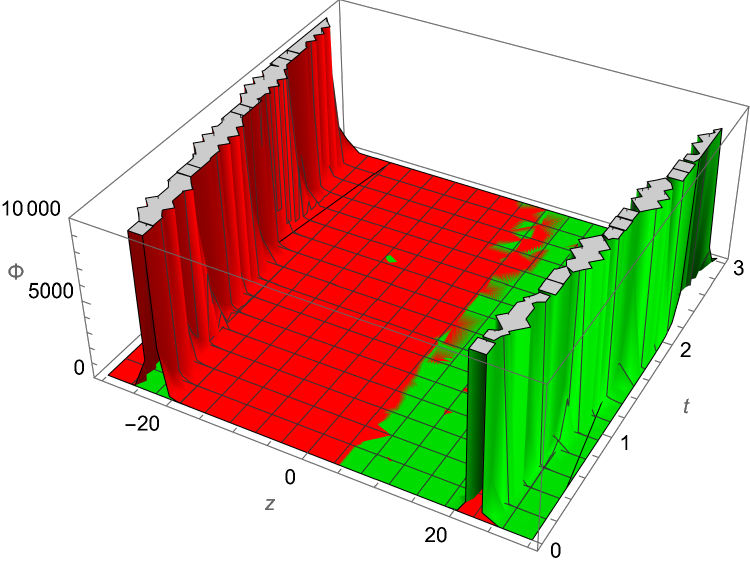}
    \caption{}
    \label{fig:first}
\end{subfigure}
\hfill
\begin{subfigure}{0.45\textwidth}
    \includegraphics[width=\textwidth]{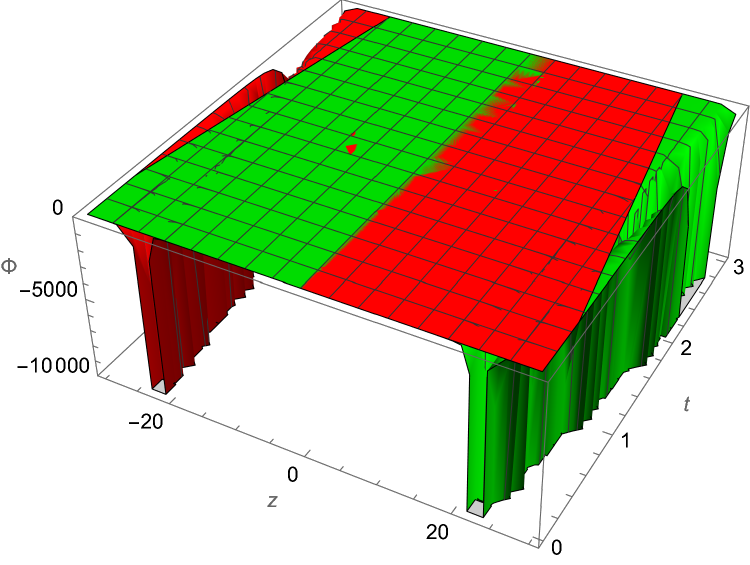}
    \caption{}
    \label{fig:second}
\end{subfigure}

\caption{(a) Plot of the Ricci components $\Phi_{22}(t,z)$ and $\Phi_{00}(t,z)$ in the $(t,z)$ coordinates satisfying the constraint $(a+b)t+(a-b)z<\frac{\pi}{\sqrt{2}}\sqrt{1-2k}$. The choice of parameters are $a=0.2$, $b=0.1$ and $k=1/4$. Note that interchanging the parameters $a\leftrightarrow b$ gives $\Phi_{22}\leftrightarrow \Phi_{00}$. (b) Plot of the Ricci component $\Phi_{02}(t,z)$ under same parameters as adapted in part (a) with $a\leftrightarrow b$.}
\label{fig:figures}
\end{figure}

\newpage

\section{The physical interpretation of our spacetime}

\subsection{Collision of pure gravitational waves}

In order to understand the present problem, we wish to review briefly the collision of gravitational waves that result locally in the Schwarzchild metric. From region II and III, we have the incoming Weyl components (see Appendix A)
\begin{equation}
\Psi_4(u)=a\delta(u)-\frac{3a^2\Theta(u)}{(1+sin(au))^3}
\end{equation}

\begin{equation}
\Psi_0(v)=b\delta(v)-\frac{3b^2\Theta(v)}{(1+sin(bv))^3}
\end{equation}

respectively, where as before $(a, b)$ are constants. Upon collision we have

\begin{equation}
\Psi_4(u, v)=\frac{a\delta(u)}{cos(bv)(1+sin(bv))^2}-\frac{3a^2\Theta(u)}{(1+sin(au+bv))^3}
\end{equation}

\begin{equation}
\Psi_0(u, v)=\Psi_4(u\leftrightarrow v, a\leftrightarrow b)
\end{equation}

\begin{equation}
\Psi_2(u, v)=\frac{ab\theta(u)\theta(v)}{cos(v)X^3}
\end{equation}

with $X=1+sin(au+bv)$. In the interaction region $u>0, v>0$, the impulsive terms are ignored and what remains satisfy the type-D condition

\begin{equation}
9\Psi_2^2=\Psi_0\Psi_4
\end{equation}

as it should. Projection of the metric into the $x=const$, $y=const$ plane gives
\begin{equation}
ds^2=2(1+sin(au+bv))dudv
\end{equation}

which is simple enough to study the Penrose diagram. For $au+bv=\frac{\pi}{2}$, we have $r=2$, which corresponds to the event horizon of the static spacetime. For $au+bv=\pi$, we have $r=1$ and for $au+bv=\frac{3\pi}{2}$, we get $r=0$, the spacelike singularity (Fig. 5a,5b). In the pure gravitational wave collision problem therefore it is seen that the entire interaction region takes place in between $0<r<2$. For more details of this particular case, we refer \cite{3}.

\subsection{Collision of null matter sources}

By introducing an incoming null source with the parameter $k\neq0$, from the theory of bumblebee gravity many features encountered in part (A) needs revision. First of all, in the incoming regions II and III, in order to define a physical source, such as $\Phi_{22}>0$, and $\Phi_{00}>0$, we must choose $k<\frac{1}{2}$. This automatically has restrictions on the acceptable range of the null coordinates u and v. By the local isometry the metric (12) obtained from the static one and the NP curvature/Ricci components (see Appendix A) suggest that we must choose $au<\frac{\pi}{2}\sqrt{1-2k}$ and $bv<\frac{\pi}{2}\sqrt{1-2k}$. As a result, in the interaction region we end up with the limitation

\begin{equation}
au+bv<\frac{\pi}{2}\sqrt{1-2k}.
\end{equation}

Now, projecting the metric to the $(u, v)$ plane gives us the same metric form as in (32). Unlike the case of colliding pure gravitational waves isometric to the Schwarzchild geometry the interaction region now is restricted by the condition $1<r<1+sin(\frac{\pi}{2}\sqrt{1-2k})$, which is applicable only for $u>0$, $v>0$ and $k\neq0$. The event horizon of the vacuum at $r=2$, is not accessible in the present case. Given the condition (33), the data for the ejected jet-like structures from the Ricci tensor components are bound to stay inside the region bounded by the 'fold' singularity and in particular in the shaded region of Fig. 5c of the Penrose diagram . It should be reminded that already the jets take place in the vicinity of black holes and any horizon to form becomes time dependent, shrinking in time. In the absence of horizon or in the presence of naked singularities, the observability of the forming jets becomes more clear.

\begin{figure}[h]
  \centering
  \subfloat[]{\includegraphics[width=0.30\textwidth]{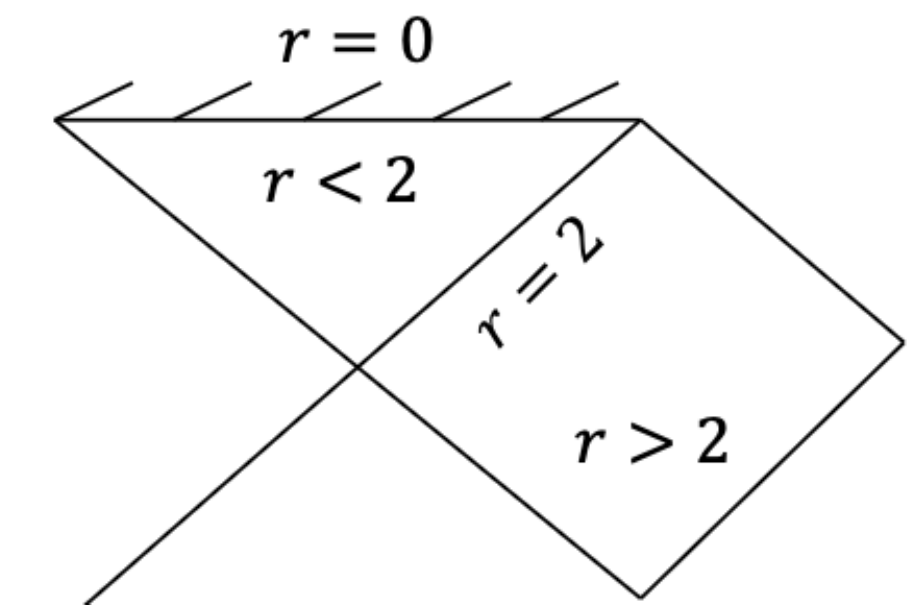}}\ 
  \subfloat[]{\includegraphics[width=0.24\textwidth]{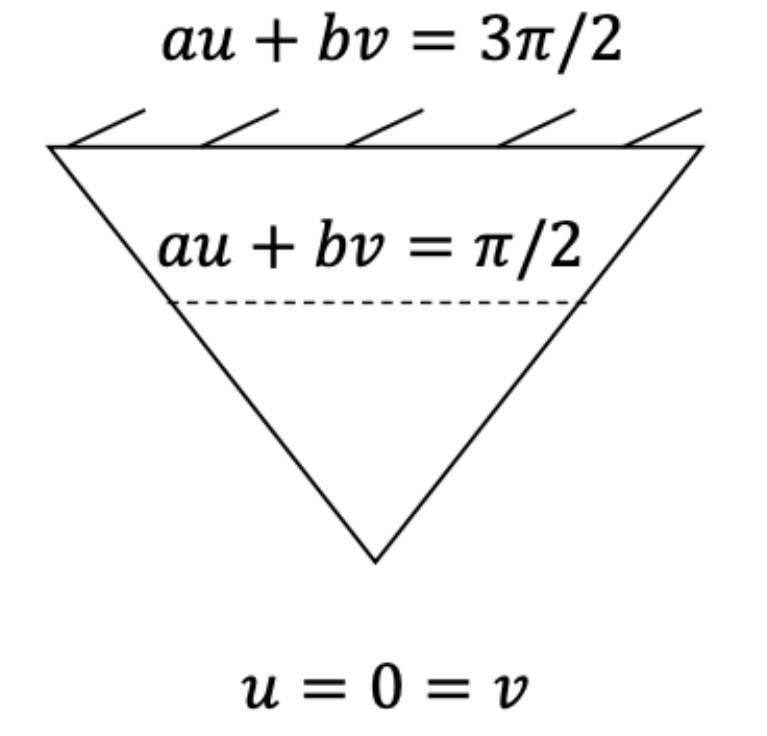}}\ 
  \subfloat[]{\includegraphics[width=0.34\textwidth]{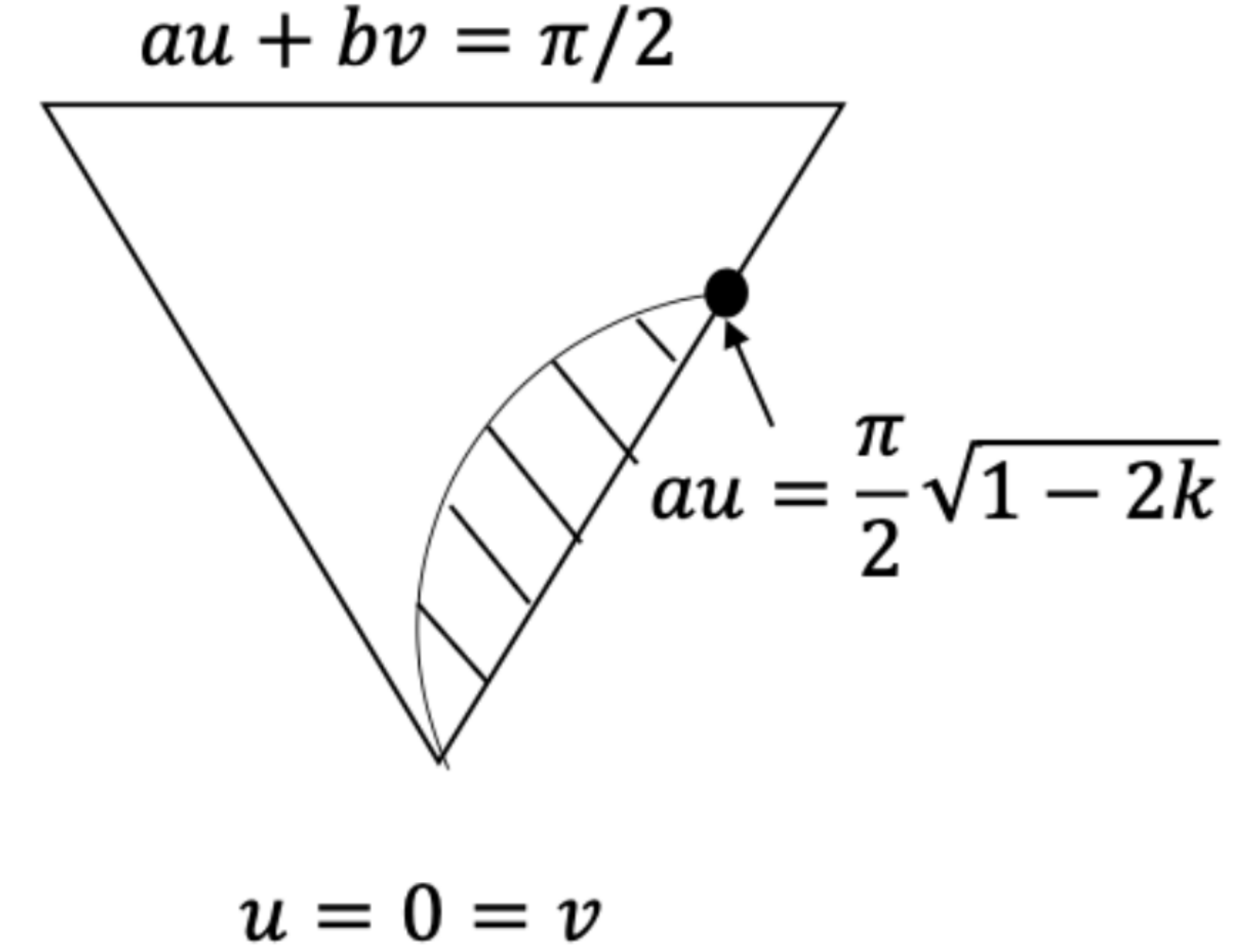}}
  \caption{(a) Part of the Penrose diagram representing the static spacetime for $0\leq r<\infty$ with $m=1$. The role of the parameter $k$ is to add a constant to the tortoise coordinate in the Schwarzchild spacetime. Our region of collision occurs in the region denoted by $0<r<2$. (b) Penrose diagram for the interaction region of colliding gravitational waves isometric to the Schwarzchild spacetime from the instant of collision $u=v=0$, $au+bv=\frac{\pi}{2}$ is a horizon where $au+bv=\frac{3\pi}{2}$ corresponds to the spacelike singularity of static spacetime $r=0$. (c) Penrose diagram for colliding null matter represented by the line element (12) in the text. The singularity is reached earlier at the indicated heavy dot point and its symmetric point with $u \leftrightarrow v$, $a \leftrightarrow b$. The reason may be attributed to the stronger focusing by the null matter relative to the pure gravitational wave problem. Geodesic particles and jet-like peaks emerge all in the shaded region and its symmetric region. Note that the previous bound for gravitational waves, i.e, $au+bv=\frac{\pi}{2}$ is not accessible in the present case.}
  \label{fig:foobar}
\end{figure}

\section{Conclusion}

Can astrophysical jets observed be described within the context of general relativity (GR)? This is the question that we addressed in this article in a restricted sense. Since jets emanate from strongly attractive sources such as black holes, our method is to consider collision of oppositely moving sources originated from the accretion disks. In such a process, the attracted sources reach almost to the speed of light, a requirement of symmetry to obtain an exact solution. The subject of colliding waves is a well-established one in GR and our intention is to understand the formation of jets through this subject. Due to assumed symmetry in going to the speed of light, we invite also undesired singularities which arise prior to the expected singularities due to gravitational waves. This implies that the null sources focus each other much stronger than the waves. The subject of singularities and their \cite{16,18} possible removal constitutes by itself a separate title which is not addressed in this study. We took these singularities for granted, excised them in favour of the regular part of the spacetime since from the physical standpoint there should be no singularity at all. The source employed is a global monopole coupled to a Schwarzchild source or equivalently either a cloud of string or a static source in the bumblebee gravity. This choice is only due to its simplicity, since it is the simplest extension of the Schwarzchild source. This should be understood in the sense of an infinitely Lorentz boosted mass into a plane fronted gravitational wave or in analogy with a point charge in electromagnetism \cite{9}. Once this is done, it is natural to study the collision of such waves+sources, moving in opposite directions. Upon collision of sources moving opposite and containing Heaviside step functions it is natural to expect that redistribution of source  takes place. Specifically, emergence of impulsive null shells due to mutually strong focusing characterized by Dirac delta function seems inevitable. It is our belief that the junction conditions on null hypersurfaces \cite{13} can be considered within the context of alternative theories to general relativity. The relation of colliding waves and black holes was discussed in  \cite{14}. It is easily seen that whenever $k=0$, no jet-like formations remain and the problem reduces to the well-known problem of colliding gravitational waves which is isometric to the Schwarzschild geometry. The essence of such a local isometry is based on the fact that inside the event horizon the static spacetime becomes dynamical. One of the spacelike coordinate becomes time coordinate and vice versa for the static metric time. As a result the line element turns into a time dependent one. In such a spacetime there are moving waves and those that are not parallel naturally collide. We employ the static line elements precisely for this purpose. It is natural to expect that beside the moving  source, boosted charges and spinning sources will add novelties to the problem. In particular, given the dynamical structure of horizons leading to observable jets from outside, in analogy with detecting the Hawking radiation from a collapsing object remains  to be explored. Admittedly our jet-like objects are visible only to the observers lying within the horizon of our static model. Our final remark is that the method employed in this paper is applicable, through the local isometry between black holes and colliding waves, to any Schwarzschild-like metric. The application extends even beyond  Schwarzchild-like metrics and applies also to the Lovelock gravity \cite{15}.

\appendix

\section{}

In the null-tetrad basis of NP, our tetrad components that are not zero are as follow

\begin{equation}
    \begin{array}{lr}
         \Lambda=\frac{R}{24}=\frac{kab\theta(u)\theta(v)}{3(1-2k)X^2}\\

         \Phi_{02}=\phi_{20}=-\frac{kab\theta(u)\theta(v)}{(1-2k)X^2}\\

         \Phi_{22}=\frac{a^2\delta(u)}{2X^2}[\tan(bv\theta(v))-\frac{1}{\sqrt{1-2k}}\tan(\frac{bv\theta(v)}{\sqrt{1-2k}})]+\frac{ka^2\theta(u)}{(1-2k)X^2}\\

         \Phi_{00}=\frac{b^2\delta(v)}{2X^2}[\tan(au\theta(u))-\frac{1}{\sqrt{1-2k}}\tan(\frac{au\theta(u)}{\sqrt{1-2k}})]+\frac{kb^2\theta(v)}{(1-2k)X^2}\\

         \Psi_2=\frac{ab\theta(u)\theta(v)}{3X^3}(\frac{kX}{1-2k}+3)\\

         \Psi_4=\frac{a\delta(u)}{2X^2}[\frac{1}{\cos(bv\theta(v))}+\frac{\cos(bv\theta(v))}{X}\\

         +\frac{1}{\sqrt{1-2k}}\tan(\frac{bv\theta(v)}{\sqrt{1-2k}})]-\frac{a^2\theta(u)}{X^3}(\frac{kX}{1-2k}+3)\\

         \Psi_0=\frac{b\delta(v)}{2X^2}[\frac{1}{\cos(au\theta(u))}+\frac{\cos(au\theta(u))}{X}\\

         +\frac{1}{\sqrt{1-2k}}\tan(\frac{au\theta(u)}{\sqrt{1-2k}})]-\frac{b^2\theta(v)}{X^3}(\frac{kX}{1-2k}+3)\\

    \end{array}
\end{equation}

where $X=1+\sin(au\theta(u)+bv\theta(v))$. Our notation is such that, $\delta(u)/\delta(v)$ and $\theta(u)/\theta(v)$ stand for Dirac delta and Heaviside step functions, respectively.

\end{document}